\documentclass[journal]{IEEEtran}

\usepackage{graphicx}
\usepackage{amsmath}
\usepackage{amssymb}
\usepackage{color}

\ifCLASSOPTIONcompsoc
  \usepackage[caption=false,font=normalsize,labelfont=sf,textfont=sf]{subfig}
\else
  \usepackage[caption=false,font=footnotesize]{subfig}
\fi

\newcommand{\vtheta}{\boldsymbol{\theta}}

\newcommand{\vx}{\boldsymbol{x}}

\newcommand{\sone}{\texttt{1}}
\newcommand{\stwo}{\texttt{2}}

\newcommand{\sfive}{\texttt{5}}

\newcommand{\vL}{\mathsf{L}}
\newcommand{\vD}{\mathsf{D}}
\newcommand{\vW}{\mathsf{W}}
\newcommand{\vI}{\mathsf{I}}

\newcommand{\RR}{\mathbb{R}}

\newcommand{\ud}{\textup{d}}

\newcommand{\vZ}{\boldsymbol{Z}}
\newcommand{\vJ}{\boldsymbol{J}}
\newcommand{\vC}{\boldsymbol{C}}
\newcommand{\vzeta}{\boldsymbol{\zeta}}

\newcommand{\sr}{\mathsf{r}}
\newcommand{\se}{\mathsf{e}}
\newcommand{\sfs}{\mathsf{s}}

\begin{document}

\title{Assess Sleep Stage by Modern Signal Processing Techniques}

\author{Hau-tieng~Wu,~Ronen~Talmon,~Yu-Lun~Lo*
\thanks{H.-T. Wu is with the Department of Mathematics, University of Toronto, Toronto Ontario Canada. (email: hauwu@math.toronto.edu)}
\thanks{R. Talmon is with the Department of Electrical Engineering, Technion - Israel Institute of Technology, Haifa, 32000, Israel. (email: ronen@ee.technion.ac.il)}
\thanks{*Correspondence: Y.-L. Lo is with the Department of thoracic medicine, Chang Gung Memorial Hospital, Chang Gung University, School of Medicine, Taipei, Taiwan. (email: loyulun@hotmail.com)}}

\markboth{Journal of \LaTeX\ Class Files,~Vol.~11, No.~4, December~2012}%
{Shell \MakeLowercase{\textit{et al.}}: Bare Demo of IEEEtran.cls for Journals}

\maketitle

\begin{abstract}

In this paper, two modern adaptive signal processing techniques, Empirical Intrinsic Geometry and Synchrosqueezing transform, are applied to quantify different dynamical features of the respiratory and electroencephalographic signals. 
We show that the proposed features are theoretically rigorously supported, as well as capture the sleep information hidden inside the signals.
The features are used as input to multiclass support vector machines with the radial basis function to automatically classify sleep stages.  
The effectiveness of the classification based on the proposed features is shown to be comparable to human expert classification -- the proposed classification of awake, REM, N1, N2 and N3 sleeping stages based on the respiratory signal (resp. respiratory and EEG signals) has the overall accuracy $81.7\%$ (resp. $89.3\%$) in the relatively normal subject group.  
In addition, by examining the combination of the respiratory signal with the electroencephalographic signal, we conclude that the respiratory signal consists of ample sleep information, which supplements to the information stored in the electroencephalographic signal.
\end{abstract}


\IEEEpeerreviewmaketitle

\section{Introduction}

\IEEEPARstart{I}{n} human beings, sleep is a universal recurring dynamical and physiological activity, and the quality of sleep influences our daily lives in diverse ways. However, it was not until recently that sleep became a brach of medicine and found its role in several seemingly unrelated clinical problems. Physiologically, it is divided into two broad stages: rapid eye movement (REM), and non-rapid eye movement (NREM) \cite{Lee-Chiong:2008}. Normally, sleep proceeds in cycles in between REM and NREM.
The NREM stage is further divided into shallow sleep (stage N1 and N2) and deep sleep (stage N3). In all procedures identifying sleep stages, we need a sleep scoring process with the help of polysomnography (PSG), which includes electroencephalography (EEG), electromyogram (EMG), and electrooculogram (EOG), etc. 

Among these physiological signals, EEG signals are the most concentrated ones since the clinically acceptable stage of the sleep is majorly determined by reading the recorded EEG based on the R\&K criteria, which were standardized in 1968 by Allan Rechtschaffen and Anthony Kales \cite{Rechtschaffen_Kales:1968} and further developed by the American Academy of Sleep Medicine on 2007 (AASM 2007) \cite{Iber_Ancoli-Isreal_Chesson_Quan:2007}. 
However, due to the subjective judgement and different training background, the agreement of manual sleep scoring among trained clinicians and professionals has been known to be limited \cite{Norman_Pal_Stewart_Walsleben_Rapoport:2000}, thereby motivating the development of an objective and automatic scoring system.

Based on these clinical findings, various features of the EEG signals have been proposed to study the sleep dynamics, for example, time domain summary statistics,  spectral analysis, coherence, time-frequency analysis, entropy, to name but a few \cite{Bajaj_Pachori:2013,Kannathal_Choo_Acharya_Sadasivan:2005,Blanco_Quiroga_Rosso_Kochen:1995,Geng_Zhou_Yuan_Cai_Zeng:2011}. Recently, a theoretically solid approach suitable to model the underlying dynamics of the brain activity and estimate the evolutionary dynamics from recorded EEG signal was proposed in \cite{TalmonPNAS,TalmonACHA}, and had been successfully applied to predict the pre-seizure state from the intra-cranial EEG signals \cite{Duncan_Talmon_Zaveri_Coifman:2013,TalmonTSP}.

However, it is well known that sleep is a global and systematic behavior not localized solely in the brain. For example, the muscular atonia and low amplitude EMG are related to the significant changes in the breathing pattern during normal sleep: during NREM sleep, especially stage N3, breathing is remarkably regular, while during REM sleep, breathing is irregular with sudden changes. 
The above physiological facts hint that the respiratory pattern of the recorded breathing signal during sleep might well reflect the sleep stage. 
There have been some reported studies of the sleep stage from analyzing the respiratory signal  \cite{Chung_Choi_Kim_Lim_Choi_Jeong_Park:2007,Guerrero-Mora_Elvia_Bianchi_Kortelainen:2012,Sloboda_Das:2011,Wu:2013,Chen_Cheng_Wu:2013}. In \cite{Chung_Choi_Kim_Lim_Choi_Jeong_Park:2007} (resp. \cite{Guerrero-Mora_Elvia_Bianchi_Kortelainen:2012}), an averaged respiratory rate over a fixed window is used to estimate the REM and NREM (resp. awake and sleep). In \cite{Sloboda_Das:2011}, a notch filter based instantaneous frequency estimator is applied to extract features to differentiate awake, REM and NREM. In \cite{Wu:2013,Chen_Cheng_Wu:2013}, the {\it adaptive harmonic model} and a modern time-frequency analysis technique have been applied to further quantify the notion of respiratory dynamic. In particular, the instantaneous respiratory rate has been related to awake, REM, shallow and deep sleep stages, with a rigorous mathematical foundation. 

The above-mentioned physiological patterns inside the EEG and the respiratory signals are actually outcomes of the intricate deformation of the underlying sleep dynamics, which we call {\it intrinsic dynamical features} of the sleep, that are not directly accessible to us. 
Although it is not an easy task to fully model or estimate the dynamical system underlying sleep, we might expect to benefit if we are able to quantify and integrate these hidden intrinsic dynamical features. In this paper, we propose to combine two modern adaptive signal processing techniques, {\it Empirical Intrinsic Geometry (EIG)} and {\it Synchrosqueezing transform (SST)}, to estimate these intrinsic dynamical features of sleep guiding the observed EEG and respiratory signals -- we define an index, referred to as {\it Sleep Index}, to quantify these features. Then, by applying the suitable classifier algorithm, we show that the extracted features are highly correlated to the sleep stage determined by reading the EEG by the AASM 2007 criteria. Indeed, the proposed classification based on the respiratory signal (resp. respiratory and EEG signals) has the overall accuracy $81.7\%$ (resp. $89.3\%$) in the relatively normal subject group, which is comparable to human expert classification. 

The article is organized in the following way. In Section \ref{Section:Method}, we summarize the theoretical background of EIG and SST and the associated models. Then the Sleep Index is introduced in Section \ref{Section:SleepIndex}. In Section \ref{Section:Result}, the proposed Sleep Index is applied to study the whole night sleep signals.
We conclude with discussion in Section \ref{Section:Discussion}.

\section{Two Algorithms -- Synchrosqueezing transform and Empirical Intrinsic Geometry}\label{Section:Method}

The work presented in this paper is an application of the modern signal processing techniques to study the sleep dynamics. In particular, we will extract different features from the respiratory and EEG signals by the well studied EIG \cite{TalmonPNAS,Talmon_Cohen_Gannot_Coifman:2013} and SST \cite{Daubechies_Lu_Wu:2011,Chen_Cheng_Wu:2013}. As such, the theoretical material will be presented in a compact, informal manner emphasizing on the intuitions. We provide a formal and rigorous summary without proof of the details. Those interested in the proofs are encouraged to read the associated references.

\subsection{Adaptive Harmonic Model and Synchrosqueezing Transform}

The major characteristic pattern of the respiratory signal is that it is almost periodic. We call the movement of air from the environment into the lungs {\em inspiration} and the movement of air in the opposite direction {\em expiration}. An inspiration and an expiration constitute {\em a respiratory cycle}. {\em Breathing process} is a physiological process consists of a sequential respiratory cycles. In this paper, we focus on the breathing process and call the time-varying volume occupying the lung space {\em the physiological respiratory signal}.   
This general observation leads us to the following {\it phenomenological model} for the respiratory signal $R(t)$ (without noise):
\begin{equation}
R(t) = A(t) \, s(\phi(t)),
\label{decomp1}
\end{equation}
where we shall call $s(\cdot)$ the wave shape function; it is a $1$-periodic real function that satisfies some mild technical conditions. See \cite{Wu:2013} for the details. The respiratory signals recorded from different devices, like the airflow measuring device or the chest wall movement, shall be understood as observations of the respiratory system. Different observations lead to different shape functions. We call the derivative $\phi'(t)$ of the function $\phi(t)$ the {\it instantaneous frequency} (IF) of the respiratory signal $R(t)$. We require IF to be positive, but it does not required to be constant as long as the variations are slight from one period to the next, i.e. $|\phi''(t)|\leq \epsilon\phi'(t)$ for all time $t$, where $\epsilon$ is some small, pre-assigned positive number. Likewise, We call $A(t)$ the {\it amplitude modulation} (AM) of $R(t)$, which should be positive, but is allowed to vary slightly as well, i.e. $|A'(t)|\leq \epsilon\phi'(t)$ for all time $t$. We refer the interested reader to \cite{Chen_Cheng_Wu:2013} for the technical details and a further discussion of the well-definedness of the definition of AM and IF. Note that our treatment of the respiratory signal is purely phenomenological; that is, the parameters and indices we will derive from the signal will be based solely on these signals themselves, and not on explicit, quantitative models of the underlying mechanisms. 

Physiologically, the quantities $\phi'(t)$, $A(t)$ and $s$ in the model (\ref{decomp1}) quantify the dynamics of the breathing process, which we refer to as {\it phenomenological dynamical features}. For example, one way to quantify the widely used notion {\it breathing rate variability (BRV)} \cite{Engoren:1998,Benchetrit:2000,Wysocki_Cracco_Teixeira_Mercat_Diehl_Lefort_Derenne_Similowski:2006} is considering the IF and AM \cite{Wu_Hseu_Bien_Kou_Daubechies:2013}. Indeed, if $\phi'(t_0)>\phi'(t_1)$ where $t_1\neq t_0$, we know that the subject breaths faster at time $t_0$ than at time $t_1$. We mention that this ``fast-slow'' momentary behavior in the respiratory signal has been shown to be clinically informative and can be helpful in the ventilator weaning prediction \cite{Wysocki_Cracco_Teixeira_Mercat_Diehl_Lefort_Derenne_Similowski:2006,Wu_Hseu_Bien_Kou_Daubechies:2013} and sleep stage estimation \cite{Wu:2013,Chen_Cheng_Wu:2013}. 

Due to the inevitable measurement error and other outliers appearing inside the system, we model the {\it recorded respiratory signal} as 
\begin{align}\label{observation_signal}
Y(t) =R(t)+\sigma(t)\xi(t),
\end{align} 
where $\xi(t)$ is a stationary generalized random process and $\sigma$ is a smooth function which varies slowly. Here $\xi(t)$ models the noise and other outliers and $\sigma$ models the possible non-stationary nature of the noise. A particular example for $\xi$ frequently encountered in practice is the Gaussian white noise. We refer the read having interest to \cite{Chen_Cheng_Wu:2013} for further information about this noise model and its mathematical details.

\begin{figure}[h]
\begin{centering} 
\includegraphics[width=.5\textwidth]{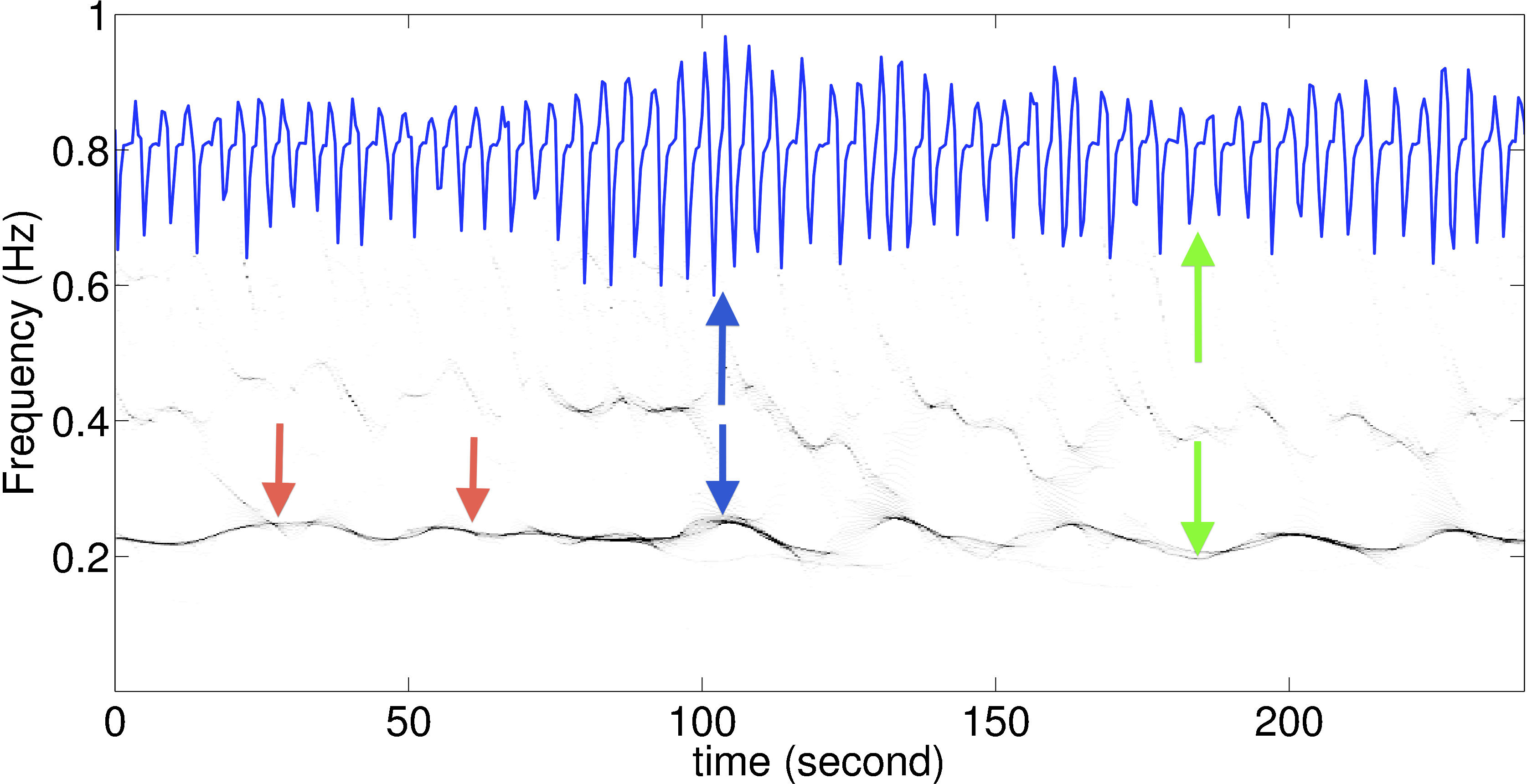} 
\end{centering}
\caption{The time-frequency representation of the respiratory signal determined by the Synchrosqueezing transform (SST) superimposed by the respiratory signal (the blue curve). The dominant black curve shown in the time-frequency representation indicated by the red arrows is the instantaneous frequency (IF) of the respiratory signal. It is clear that the subject breathes faster during the time indicated by the blue arrow than that indicated by the green arrow. This observation is captured by the IF indicated by the blue and green arrows.}
\label{fig:1}
\end{figure}

To estimate the phenomenological dynamical features of $R(t)$, $\phi'(t)$ and $A(t)$, from $Y(t)$, we apply the {\it Synchrosqueezing transform} (SST), which is a special reallocation technique \cite{Daubechies_Lu_Wu:2011,Chen_Cheng_Wu:2013}. In a nutshell, we evaluate any linear time-frequency analysis on the observation $Y(t)$, for example the short time Fourier transform or the continuous wavelet transform, and we take the phase information hidden inside the chosen linear time-frequency analysis into account to obtain a sharpened time-frequency representation, which is denoted as $S_Y(t,\xi)$. In addition to capturing the phenomenological dynamical features of $R(t)$, the SST provides a sharper time-frequency representation compared with the other traditional time-frequency analyses.  
See Figure \ref{fig:1} for an example of the respiratory signal and its instantaneous frequency. We refer the reader to \cite{Daubechies_Lu_Wu:2011,Chen_Cheng_Wu:2013} for more details. 

\subsection{Empirical Intrinsic Geometry and its underlying Mathematical Model}
\label{Subsec:EIG}

In many real-world applications, the seeming complicated time series collected from the system is controlled by a relatively simple underlying process. In some situations, when the underlying evolutionary process lies on a low-dimensional Riemannian manifold, it can be parameterized through a manifold learning framework, which was first introduced and studied in \cite{Coifman_Singer:2008}. 
The main idea in \cite{Coifman_Singer:2008}\footnote{In the original paper \cite{Coifman_Singer:2008}, the method was referred to as Nonlinear Independent Component Analysis. However, for the sake of avoiding possible confusion with Independent Component Analysis, the name Empirical Intrinsic Geometry (EIG) was adpated in \cite{TalmonPNAS,TalmonACHA}.} and its extensions to time-series \cite{TalmonPNAS,TalmonACHA} is to bridge between data mining, and in particular manifold learning, and dynamical systems. The authors' observation that the accessible measurements at hand do not necessarily convey the true essence of the system led to the development of a more generalized model, which separates between measurements and intrinsic hidden variables.

One particular example for such a dynamical system is the respiratory signal recorded during sleep -- we consider the model that the evolutionary process governing the respiratory signals is restricted to a low-dimensional Riemannian manifold, which is fundamentally different from the phenomenological model \eqref{observation_signal}. This dependency is encoded using the {\it state-space} formalism and the model of the recorded respiratory signal \eqref{observation_signal} is extended as follows: 
\begin{equation}\label{state_space}
\left\{
\begin{array}{ll}
Y(t) =R_{\vtheta}(t)+\sigma(t)\xi(t),&\,\, \mbox{[measurement equation]}\\
\ud\theta_i(t)=a_i(\theta_i(t))\ud t+\ud w_i(t),&\,\, \mbox{[state equation]}
\end{array}
\right.
\end{equation}
where $\vtheta(t):=(\theta_1(t),\ldots,\theta_d(t))$ forms the {\it inaccessible intrinsic state} at time $t$ that governs the respiratory signal $R_\theta(t)$ and evolves in time with unknown drifts $a_i$ and independent standard Brownian motions $w_i$, $i=1,\ldots,d$.

The idea that lies behind the model \eqref{state_space} is twofold. First, the measured signal $Y(t)$ has typical (unknown) dynamics, modeled here by the state equation, which need to be taken into account and encoded in the desired features. Second, the accessible signal is viewed as a measurement of the neural system controlling the sleep cycle. While it can be effected by numerous factors relating to the measurement modality (e.g., measurements of airflow or chest movements), the used equipment (e.g., the type of sensors and their exact positions), and noise, the true intrinsic variable we have interest in is the intrinsic states controlling the respiratory signal (represented here by $\vtheta (t)$). Indeed, the notation $R_{\vtheta}(t)$ implies that the respiratory signal depends upon the ``real" physiological variable $\vtheta$ in an unknown way, possibly through its amplitude $A _\theta (t)$, wave shape $s _\theta (\cdot)$, or instantaneous frequency $\phi' _\theta (t)$.

In the above model \eqref{state_space}, however, due to noise and other nuisance factors, the measurement $Y(t)$ might be too redundant to faithfully describe the dependency of the the respiratory signal on the underlying state and its temporal evolutionary. Thus, to improve the underlying state observability, we introduce a {\it high dimensional (possibly nonlinear) observer} $\Phi$ to the measured signal \cite{TalmonTSP}, i.e.,
\begin{equation}\label{equation:observation}
	\vZ(t) = \Phi(Y(t)) \in \mathbb{R}^m, \quad \mbox{[observation equation]}
\end{equation}
where $\Phi$ is a map from the suitable scalar valued functional space to the $\RR^m$-valued functional space and $m\geq1$ is an integer specified by the observer. 

With the sampled observation set $\mathcal{Z}:=\{\vZ(t_i)\}_{i=1}^N$, a natural question is how to estimate $\vtheta(t)$, namely, the system intrinsic state and dynamics of interest. Such an analysis may complement the phenomenological dynamical features provided by the SST. While the SST mainly carries instantaneous information, recovering the intrinsic state of the dynamical system $\vtheta (t)$ provides a characterization of coarser, slower dynamical changes of the shape and structure of the signal, especially when the observer $\Phi$ is implemented as a transform that relies on short time frames analysis.

It was shown in \cite{Coifman_Singer:2008} that if the observations $\vZ (t)$ can be written as a regular deterministic function $f:\mathbb{R}^d \rightarrow \mathbb{R}^m$ of the samples of the underlying state, i.e., $\vZ(t) = f(\vtheta (t))$, then, by It\^o's formula, we have
\begin{equation}
\ud Z_j(t)=\sum_{i=1}^d\left(\frac{1}{2}f^j_{ii}+a_i f^j_i\right)\ud t + \sum_{i=1}^d f^j_i\ud w_i(t)
\end{equation}
where $f^j_i=\partial f_j / \partial \theta _i$ and $f^j_{ii} = \partial ^2 f_j / \partial \theta _i ^2$.
By a direct calculation, the covariance matrix $\vC(t) \in \RR^{m\times m}$ of the observation at time $t$ define by
\begin{equation}
C_{j,k}(t):=\text{Cov}(\ud y_j(t),\ud y_k(t)),
\end{equation}
satisfies $\vC(t) =\vJ (t) \vJ^T(t)$, where $\vJ=\nabla f \in \RR^{m\times d}$ is the Jacobian of $f$. This key result, along with the assumption that $\vtheta (t)$ is locally stationary evolving much more slowly than the observation scale so that the it stays closely on a low-dimensional manifold $\mathcal{M}$ embedded in $\RR^d$, which is referred to as the {\it intrinsic state manifold}, as well as the assumption that $f$ is stably invertible on its range, allow the authors in \cite{Coifman_Singer:2008} to estimate the inaccessible state through the solution of an eigenvector problem, which will be described later in this section. The main step leading to the solution theory is the following estimation. Suppose $\vtheta(t),\vtheta(\tau) \in \mathcal{M}$, $\vZ(t)=f(\vtheta(t))$ and $\vZ(\tau)=f(\vtheta(\tau))$. By the Taylor expansion of $f$, up to the error term $O(\|\vZ(t)-\vZ(\tau)\|^4)$ \cite{Coifman_Singer:2008}, we have:
\begin{equation}\label{estimation:Mdist}
\begin{array}{l}
\|\vtheta (t) - \vtheta (\tau) \|_{\RR^d}^2 =\frac{1}{2}(\vZ (t) - \vZ(\tau))^T \\
\quad \quad \quad \quad \quad \quad \times [\vC^{-1}(t)+\vC^{-1}(\tau)](\vZ (t) - \vZ(\tau)).
\end{array}
\end{equation} 
Note that in our example, the function $f$ leading to the observation depends on the observer $\Phi$. As a result, with the estimated covariance matrix from the accessible collected data $\mathcal{Z}$, we can build a {\it graph Laplacian} associated with the intrinsic state manifold from the finite observations $\mathcal{Z}$ using the estimated Euclidean distance between the corresponding underlying samples $\vtheta (t_i)$ (\ref{estimation:Mdist}). This graph Laplacian gives rise to re-parametrization of the intrinsic manifold through 
diffusion maps (DM) \cite{Coifman_Lafon:2006}. This re-parametrization procedure aiming to extract the intrinsic dynamics of the observation is referred to as Empirical Intrinsic Geometry (EIG). 

The remaining key question is the choice or design of a ``proper" observer $\Phi$ in (\ref{equation:observation}) to the system. In particular, in order to accommodate the inevitable noise in real-world signals, estimates of the conditional probability density $p(\vZ | \vtheta)$ (e.g., histograms) were proposed as observers in \cite{TalmonPNAS}. The analysis relies on the following facts: (a) any measurement noise is translated to a linear transformation in the conditional densities domain, and (b) the distance (\ref{estimation:Mdist}), which is the {\it Mahalanobis distance}, is invariant under linear transformations. Indeed, these two facts allow for the estimation of the distance between two nearby samples on the intrinsic manifold in adverse noisy conditions. However, estimating the conditional probability densities requires a large amount of data and often is not feasible. Unfortunately, standard representations based on the Fourier transform are also inadequate for respiratory signals. By linear approximation of the function $\phi(t)$ around a nearby sample at $t_0$, the respiratory signal in \eqref{decomp1} can be approximated by 
\begin{equation}
R(t) \approx A(t_0) \, s(\phi(t_0) - t_0 + \phi'(t_0)t),
\end{equation}
As a result, the modulus of the Fourier transform of $R(t)$ around $t_0$ is approximated by
\begin{equation}\label{time_deformation}
|\hat{R}(t_0, \omega) | \approx |A(t_0) \, \hat{s}(\omega / \phi'(t_0))|, 
\end{equation}
where $\omega$ is the frequency, $\hat{R}(t_0, \omega)$ is the Fourier transforms of $R(t)$ around $t_0$ and $\hat{s}(\omega)$ is the Fourier transform of $s(t)$, respectively, assuming $A(t)$ changes slowly with time. The approximation in \eqref{time_deformation} implies that even an almost linear function $\phi(t)$ (i.e., when the IF is $\phi'(t) \approx 1$) is translated to large deformations in the Fourier domain in high frequencies \cite{Mallat2012}. Consequently, if we take the short time Fourier transform as the observer, the observations exhibit instabilities, thereby leading to irregular $f$ and a poor estimation of the Euclidean distance on the underlying intrinsic state manifold \eqref{estimation:Mdist}.

To overcome the instability of the Fourier representation, following \cite{TalmonTSP}, we use the scattering transform as an observer. The scattering transform has a low variance because it is based on first order moments of contractive operators, it linearizes deformations, and it can represent effectively intermittent behavior \cite{Mallat2012,Bruna2013}.
The scattering transform is computed based on a cascade of wavelet transforms and nonlinear modulus operators \cite{Mallat2012}. Here, we briefly review the construction procedure of its first and second order levels, since they were empirically shown to provide a sufficient representation of the signals considered in this paper.

Let $\psi (t)$ be a complex wavelet, whose real and imaginary parts are orthogonal and have the same $L_2$ norm. Let $\psi _j (t)$ denote the dilated wavelet, defined as $\psi _j (t) := 2^{-j} \psi (2^{-j}t), \ \forall j \in \mathbb{Z}$.
Let $\Phi _s (R_{\vtheta}(t))$ denote the observations computed by applying the first and second level scattering transform to the signal samples $R_{\vtheta}(t)$, which are given by
\begin{align*}
	&\Phi _s (R_{\vtheta}(t)) = (|| R_{\vtheta}(t) * \psi _{j_1} (t) | * \psi _{j_2}(t)| * w (t) \\ 
	&\qquad\qquad\qquad \forall (j_1, j_2) \in \mathbb{Z}^n,  n \in \{1,2\})_{j_1,j_2}
\end{align*}
where $w(t)$ is a smoothing window, i.e., a scaling function associated with the mother wavelet.
The scattering transform has been shown to be an observer that is especially suitable for deformations and intermittencies \cite{TalmonTSP}. In particular, it was shown that it is regular with respect to time deformations. Therefore, the application of the scattering transform to the respiratory signal is particularly suitable.

Building on the generality of the described analysis, in this study, we use it to represent the EEG signals as well. As the respiratory signal, the EEG signal measures a physiological phenomenon (``the brain activity"), but, it is subject to noise, interferences, and nuisance factors. Likewise, it can be represented using a state-space model, similar to \eqref{state_space}, given by
\begin{equation*}
\left\{
\begin{array}{ll}
X(t) = E_\zeta(t) + V(t)\,&\,\, \mbox{[measurement equation]}\\
\ud\zeta_i(t)=\alpha_i(\zeta_i(t))\ud t+\ud u_i(t),&\,\, \mbox{[state equation]}
\end{array}
\right.
\end{equation*}
where $E_{\zeta}(t)\in\RR^{m'}$ and $X(t)\in\RR^{m'}$ are the clean and noisy EEG signals, $V(t)$ is a measurement noise, and $\vzeta(t):=(\zeta_1(t),\ldots,\zeta_{d'}(t))$ denotes the inaccessible intrinsic state representing the brain activity that governs the EEG signal $E_\zeta(t)$ and evolves in time with unknown drifts $\alpha_i$ and independent standard Brownian motions $u_i$, $i=1,\ldots,d'$. By applying EIG to the recorded EEG signals, we may reconstruct the intrinsic states $\vzeta$. We remark that this approach was applied to identify the pre-seizure state from intracranial EEG data \cite{Duncan_Talmon_Zaveri_Coifman:2013,TalmonTSP}. We refer the interested reader to \cite{TalmonPNAS,TalmonACHA,TalmonTSP} for more technical details and references. 

Before closing this section, we summarize the construction of the graph Laplacian parametrization. In a nutshell, the main ingredient is integrating local similarities at different scales, which leads to a global description of the data set. Unlike linear methods such as principal component analysis (PCA), a graph Laplacian parametrization embodies nonlinear relationships among the variables. In addition to the mathematical analysis results \cite{Coifman_Lafon:2006,Belkin_Niyogi:2007,Singer_Wu:2013}, it has been shown to be robust to noise perturbation \cite{ElKaroui:2010a,ElKaroui_Wu:2014} and it is computationally efficient. We outline the algorithm here and refer the readers to these literatures for further theoretical details. 

\begin{figure*}[t]
\begin{centering} 
\includegraphics[width=.5\textwidth]{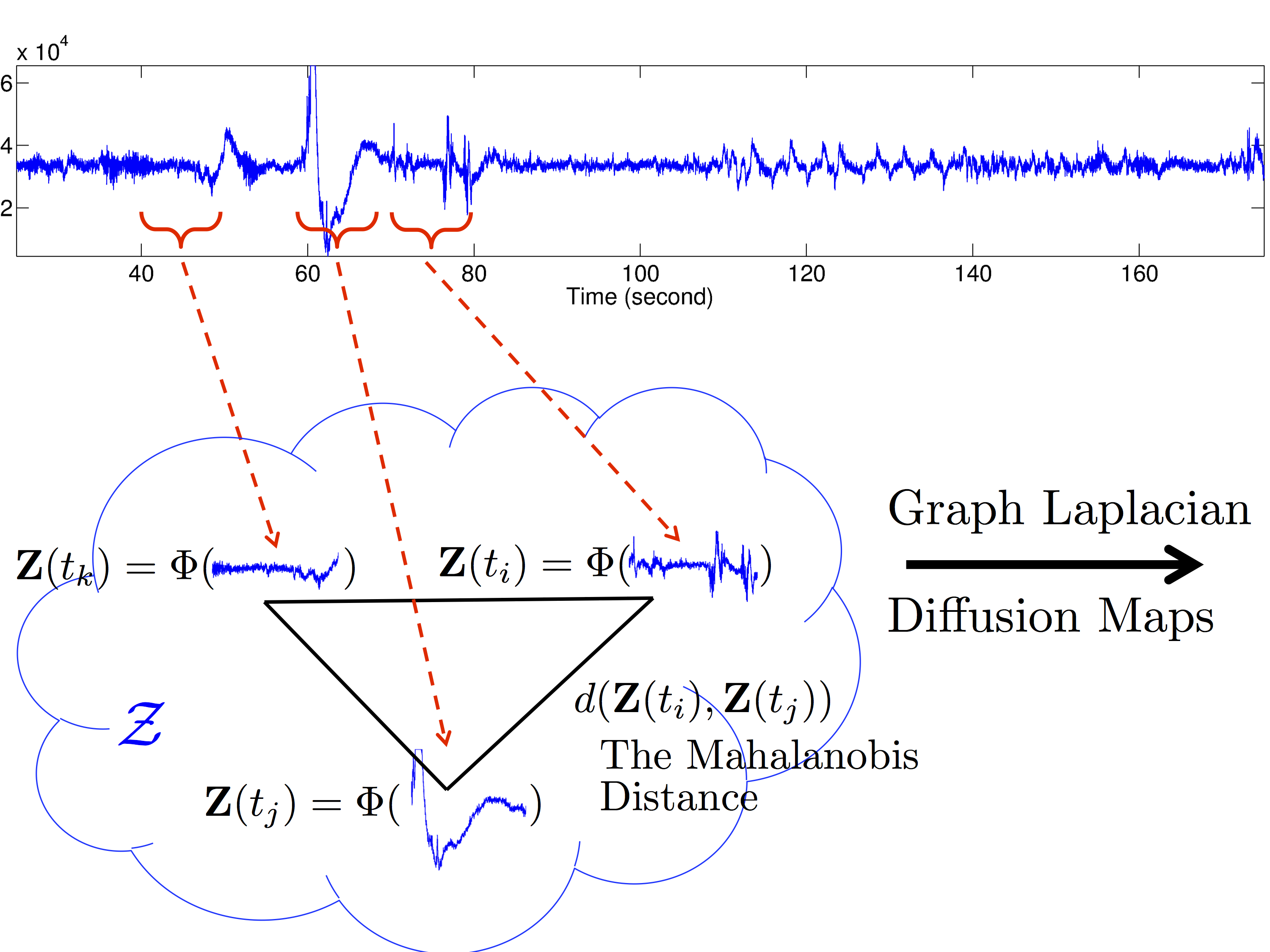}
\includegraphics[width=.5\textwidth]{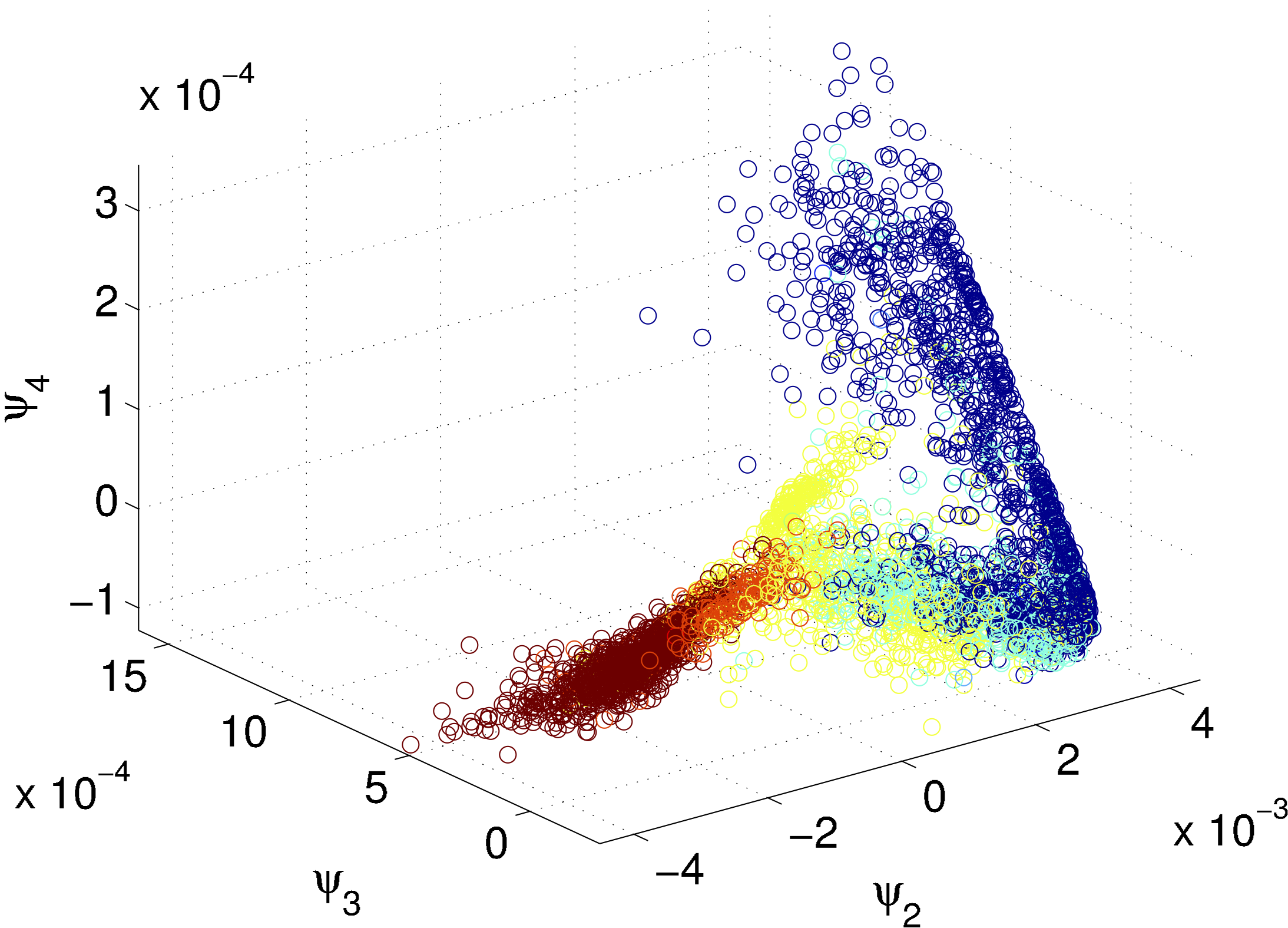} 
\end{centering}
\caption{The intrinsic dynamical features of the cortical activity extracted from the O1A2 EEG signal by the scattering Empirical Intrinsic Geometry (EIG). On the left, the scattering EIG is illustrated --  the graph Laplacian is built up from the Mahalanobis distance from the EEG signal via the scattering operator. On the right, the top three nontrivial eigenvalues of the graph Laplacian are used to show the underlying evolutionary dynamics. The blue circles (resp. cyanid circles, yellow circles and red circles) represent the awake (resp. REM, N1 and N2 and N3) sleep stage. It is clear that the extracted dynamical features well parametrize the sleep stages in the sense that different sleep stages are located in different places.}
\label{fig:2}
\end{figure*}

Take $N$ multivariate measurement samples $\mathcal{Z}=\{\vZ(t_i)\}_{i=1}^N$ 
and build a complete graph with vertices $\mathcal{Z}$. We first build an affinity matrix (or adjacency matrix) $\vW$ of size $N \times N$. The affinity between a pair of samples is defined by a metric $d$ in the following way:
\begin{equation}
\label{W}
W_{ij}=e^{-\frac{d^2(\vZ(t_i),\vZ(t_j))}{\epsilon}}, \quad \mbox{for } i,j=1,\ldots,N,\, i\neq j.
\end{equation}
Note that according to the noise analysis in \cite{ElKaroui_Wu:2014}, when the signal to noise ratio is small, it is beneficial to set the diagonal terms of the affinity matrix to $0$. 
In the present work, following the analysis in \cite{Coifman_Singer:2008}, the metric we choose is the Mahalanobis distance \eqref{estimation:Mdist}. It is clear that the matrix $\vW$ is symmetric. Note that theoretically (and practically) we can choose a more general kernel function, but we focus on the Gaussian kernel to simplify the exposition.
Then we define the diagonal degree/density matrix $\vD$ of size $n\times n$, consisting of the sum of rows of $\vW$:
$$
D_{ii}=\sum^N_{j=1}W_{ij},\quad \mbox{for } i=1,\ldots,N.
$$
Based on $\vW$ and $\vD$, the {\it graph Laplacian} is defined by
$$
\vL:=\vI-\vD^{-1} \vW.
$$
Note that under the manifold assumption, $\vD^{-1}$ exists.
Also note that $\vD^{-1}\vW$ can be viewed as a transition matrix of a Markov chain on the samples.  
Since $\vL$ is similar to the symmetric matrix $\vI- \vD^{-1/2} \vW \vD^{-1/2}$, it has a complete set of right eigenvectors $\varphi_1,\varphi_2,\ldots,\varphi_{N}$ with corresponding eigenvalues $0=\lambda_1<\lambda_2\leq\dots\leq\lambda_{N}\leq1$, where $\varphi_1=(1,1\dots,1)^T$ \cite{Coifman_Lafon:2006}. By the above construction, the eigenvectors $\varphi_1,\ldots,\varphi_{N}$ are vectors in $\mathbb{R}^N$. Through the eigenvectors, the measurement samples are mapped into $\mathbb{R}^{\hat{d}}$ via
\begin{equation}
\label{d-map}
\vZ(t_i) \mapsto (\varphi_2(t_i),\ldots,\varphi_{\hat{d}}(t_i)),\quad \mbox{for } i=1,\ldots,N.
\end{equation}
where $\hat{d}$ is an estimate of the dimension of the intrinsic state of the system and is usually $\hat{d}\ll N$. Estimating the intrinsic dimension of the system $d$ extends the scope of the paper and is empirically set according to the spectral gap in the decay of the eigenvalues, as will be described in Section \ref{Section:SleepIndex}. In \eqref{d-map}, we obtain a $\hat{d}$-dimensional parameterization of the measurements. In particular, we view the $j$th coordinate of the parameterization of $\vZ (t_i)$, i.e., $\varphi_{j+1} (t_i)$, as the $j$th coordinate of the recovered hidden intrinsic state $\theta _j (t_i)$, which we view as the features associated with the sleep stage in this analysis. An illustration of the DM reparametrization process with the first $3$ non-trivial eigenvectors is shown in Figure \ref{fig:2}.

\section{Material and Method}\label{Section:SleepIndex}

\subsection{Data Collection}

Standard polysomnography was performed with at least 6 hours of sleep to confirm the presence or absence of OSA from the clinical subjects suspicious of sleep apnea in the sleep center at Chang Gung Memorial Hospital (CGMH), Linkou, Taoyuan, Taiwan. The institutional review board of the CGMH approved the study protocol (No. 101-4968A3) and the enrolled subjects provided written informed consent. Four channel EEG signals (C3A2, C4A1, O1A2 and O2A1), two channel EOG signals and chin EMG 
were recorded at the sampling rate $200$ Hz for sleep staging. Chest and abdominal motions are recorded by the piezo-electric bands and airflow was measured using thermistors and nasal pressure, both at the sampling rate $100$ Hz. 
All signals were acquired on the Alice 5 data acquisition system (Philips Respironics, Murrysville, PA). Apneas and hypopneas were defined using AASM 2007 guidelines \cite{Iber_Ancoli-Isreal_Chesson_Quan:2007}, and the apnea-hyponea index (AHI) provided is the value determined during sleep.

Take the recorded EEG signals, denoted as $E_k$, $k=1,\ldots,4$, and the respiratory signal, denoted as $R$. Suppose the recording time period is $\mathcal{T}=[0,T]$. We divide $\mathcal{T}$ into contiguous subintervals $\mathcal{T}_i$ of $\tau$ seconds long, $i=1,\ldots,N$; that is, $\mathcal{T}=\cup_{i=1}^N \mathcal{T}_i$ and $\mathcal{T}_i\cap \mathcal{T}_j=\emptyset$ for all $i\neq j$. We call $\mathcal{T}_i$ the {\em $i$-th epoch}. We will extract $p>0$ features out of the recorded respiratory and EEG signals for each epoch.

\subsection{Features from the respiratory signal}

Given a recorded respiratory signal $R(t)$, we extract its {\it phenomenological dynamical features}, including the instantaneous frequency $\phi'(t)$ and the amplitude modulation $A(t)$ by applying the SST. Denote the estimated instantaneous frequency by $\tilde{\phi}'(t)$ and the amplitude modulation by $\tilde{A}(t)$. The mean of $\tilde{A}(t)$ restricted to the $i$-th epoch, denoted as the $A_i$, and the mean of $\tilde{\phi}'(t)$ restricted to the $i$-th epoch, denoted as $\phi'_i$, form the first two features for the respiratory signal for the $i$-th subinterval. The third feature, denoted as $v_i$, is obtained by evaluating the standard deviation of $\tilde{\phi}'(t)$ on the interval of length $30$ seconds centered on the middle of the $i$-th epoch.

We apply the analysis described in Section \ref{Subsec:EIG} to $R(t)$ in order to complement the phenomenological dynamical features and to obtain a characterization of the structural, slower underlying variables of the data. Here as well we obtain the graph Laplacian $\vL^{(R)}\in \RR^{N\times N}$. Then, the eigenvectors and eigenvalues of $\vL^{(R)}$ are given by $\vL^{(R)} \varphi_{j}^{(R)}=\lambda_{j}^{(R)}\varphi_{j}^{(R)}$. The first $\hat{d}^{(R)} \geq 1$ nontrivial eigenvectors are chosen based on the following ``spectral gap" thresholding criteria
\begin{align}\label{equation:thesholding}
\lambda_{\hat{d}^{(R)}+1 }^{(R)} <\delta \quad\mbox{ and }\quad 
\lambda_{\hat{d}^{(R)} +2}^{(R)} \geq\delta,
\end{align} 
where $0<\delta<1$ is the threshold chosen by the user. 
Thus, using \eqref{d-map}, we obtain $\hat{d}^{(R)}$ {\it intrinsic dynamical features} of the respiratory system.
 
\subsection{Features from the EEG signal}
Given the EEG signal $E_k(t)$ recorded from the $k$-th channel, we run the analysis described in Section \ref{Subsec:EIG} and obtain the graph Laplacian $\vL^{(E,k)}  \in \RR^{N\times N}$.  Then, the eigenvectors and eigenvalues of $\vL^{(E,k)} $ are given by $\vL^{(E,k)} \varphi_{j}^{(E,k)}=\lambda_{j}^{(E,k)} \varphi_{j}^{(E,k)}$ with $0=\lambda_{1}^{(E,k)}\leq\lambda_{2}^{(E,k)}\leq\ldots$. 
The first $\hat{d}^{(E,k)}\geq 1$ nontrivial eigenvectors are chosen based on the thresholding criteria (\ref{equation:thesholding}) with the same $\delta$. Using the eigenvectors, each subinterval of the EEG signal $E_k(t)$ is mapped into a sub-vector of $\hat{d}^{(E,k)}$ dimensions according to \eqref{d-map}.
By collecting the low dimensional vectors of all the channel, we obtain a vector consisting of $\sum_{k=1}^4 \hat{d}^{(E,k)}$ {\it intrinsic dynamical features of the cortical activity} for each subinterval.

\subsection{Sleep Index}
We consider the following two feature vectors. The first one is extracted only from the respiratory signal and is referred as the {\it Respiratory Index}:
\begin{align*}
\sr_i:=\big(A_i,\phi'_i,v_i,\varphi^{(R)}_{2}(i),\ldots,\varphi^{(R)}_{\hat{d}^{(R)}+1}(i)\big).
\end{align*}
The second one is extracted only from the EEG signals and is referred as the {\it EEG Index}:
\[
\se_i:=\big(\varphi_{2}^{(E,1)}(i),\ldots,\varphi^{(E,1)}_{\hat{d}^{(E,1)}+1 }(i),\ldots,\varphi^{(E,4)}_{2}(i),\varphi^{(E,4)}_{\hat{d}^{(E,4)}+1 }(i)\big). 
\] 
An analysis result of the O1A2 EEG signal, denoted as $\big(\varphi_{2}^{(E,1)}(i),\ldots,\varphi^{(E,1)}_{4}(i)\big)$, is shown in Figure \ref{fig:2}. Clearly different sleep stages represented in different colors have different features and are well clustered. In addition, these different sleep stages are organized in a continuous but nonlinear way -- from the right hand side of the figure to the left hand side we have awake, REM, N1 and N2 and deep sleep stages. 

Next, the $3$ phenomenological respiratory features, the $\hat{d}^{(R)}$ intrinsic respiratory features and the intrinsic dynamical features of the cortical activity at the $i$-th epoch are combined together to comprise the {\it Sleep Index} with $p=\sum_{k=1}^4\hat{d}^{(E,k)}+3+\hat{d}^{(R)}$:  
\[
\sfs_i:=\big(\sr_i,\se_i\big). 
\]

\subsection{Sleep Stage Classifier}

Support vector machine (SVM) is a commonly used technique for the purpose of classification in statistical learning theory \cite{Scholkopf_Smola:2002}. In a nutshell, SVM determines a hyperplane in the space separating the data set into two disjoint subsets, such that each subset is lying in a different side of the hyperplane. With the help of the reproducing kernel Hilbert space theory, SVM is generalized to the {\it kernel SVM}, which allows for classification with nonlinear relationship; that is, a nonlinear surface separating the data set into two disjoints subsets may be used. We refer the interested reader to \cite{Scholkopf_Smola:2002} for technical details.  
For the sake of identifying the (possible) nonlinear relationship between different sleep stages, in this work we choose the radial based function (RBF), $K(\vx,\vx')=\exp(-\frac{\|\vx-\vx'\|^2_2}{2\sigma^2})$, where $\sigma>0$, as the kernel function. Note that our dataset is multi-class -- the response has more than $2$ categories -- therefore, we need to further generalize the kernel SVM to the multi-class SVM to complete our mission. To this end, we apply the one versus all (OVA) classification scheme \cite{Rifkin_Klautau:2004}. Despite its simplicity, the OVA classification scheme is highly effective and useful, as was extensively shown and discussed in \cite{Rifkin_Klautau:2004}. Group data will be reported as mean $\pm$ standard deviation unless otherwise specified.

\section{Result}\label{Section:Result}

Ten subjects without sleep apnea (AHI less than $5$) were chosen for this study. The demographic characteristics of the individuals whose data was used are as follows: $6$ males and $4$ females, age: $45.9 \pm 12.3$ years, range $28-61$ years; BMI: $23.6\pm 1.9 \text{kg/m}^2$, range $21.5-28 \text{kg/m}^2$ ; AHI: $1.9 \pm 1.1$, range $0.4-3.4$. The total recorded time are of length $384\pm 27.8$ minutes with range $363-443$ minutes and we have a sleep period time of $367\pm 27.5$ minutes with range $338-428$ minutes for the sleep stage estimation.

We divide the whole night sleep into contiguous epochs of length $2.56$ seconds. We take $\delta=0.01$ and the dimension of the Sleep Index $p$ is $11.2\pm1.69$.  
We consider the sleep stages in this study: 
\begin{align*} 
\mathcal{R} &=\{\text{Awake},\,\text{REM},\,\text{N1},\text{N2},\,\text{N3}\}=:\{\sone,\stwo,\ldots,\sfive\}.
\end{align*}
Here to simplify the notation, we reindex the set of sleep stages and use the teletype-font to avoid confusion; that is, $\sone$ is the awake stage, etc.
Then we generate the different indices, $\{\sfs_i\}_{i=1}^N$, $\{\sr_i\}_{i=1}^N$ and $\{\se_i\}_{i=1}^N$, from the recorded EEG and respiratory signals. The sleep stages in $\mathcal{R}$ are determined by the sleep expert as the ground truth.
 
The OVA kernel SVM with the RBF kernel with $\sigma=1$ is applied to classify the different sleep stages. Suppose there are $n_\ell$ subintervals with sleep stage $\ell$, where $\ell=\sone,\ldots,\sfive$, in the validation dataset. Denote $n_{i,j}$ to be the number of subintervals with the sleep stage $\texttt{i}$ as the gold standard, but classified as the sleep stage $\texttt{j}$. We call the $5\times 5$ matrix $N$ with the $(i,j)$-th entry $n_{i,j}$ the {\it confusion matrix}. We also define the {\it confusion percentage matrix} $P$ as a $5\times 5$ matrix with its $(i,j)$ entry $\frac{n_{i,j}}{\sum_{j=1}^5n_{i,j}}$. We will call $P_{i,i}$ the {\it sensitivity (SE)} of the sleep stage $\texttt{i}$ prediction, which is denoted as SE($\texttt{i}$). 
We will also report the overall accuracy (AC) denoted as $\text{AC} :=\frac{\sum_{\texttt{i}=\texttt{1}}^{\texttt{5}}n_{\texttt{i},\texttt{i}}}{\sum_{\texttt{i,j}=\texttt{1}}^{\texttt{5}}n_{\texttt{i},\texttt{j}}}$ and the specificity (SP) of the sleep stage $\texttt{i}$ denoted as $\text{SP}(\texttt{i}):=\frac{ \sum_{\texttt{j}\neq \texttt{i}}n_{\texttt{j},\texttt{j}}}{\sum_{\texttt{k}}\sum_{\texttt{j}\neq \texttt{i}}n_{\texttt{j},\texttt{k}}}$. Note that these definitions are direct generalizations of the AC, SE and SP of the binary categorical response data.

To prevent over-fitting and confirm the classification result, we run the repeated random sub-sampling validation $25$ times and evaluate the average. To be more precise, we randomly partition the data into the training dataset and the validation dataset -- the training dataset  comprises $80\%$ of the features and the rest are used to form the validation dataset. The trained classifier based on the training dataset is applied to predict the sleep stages of the validation dataset.

\begin{figure*}[t]
\begin{centering}
\includegraphics[width=.7\textwidth]{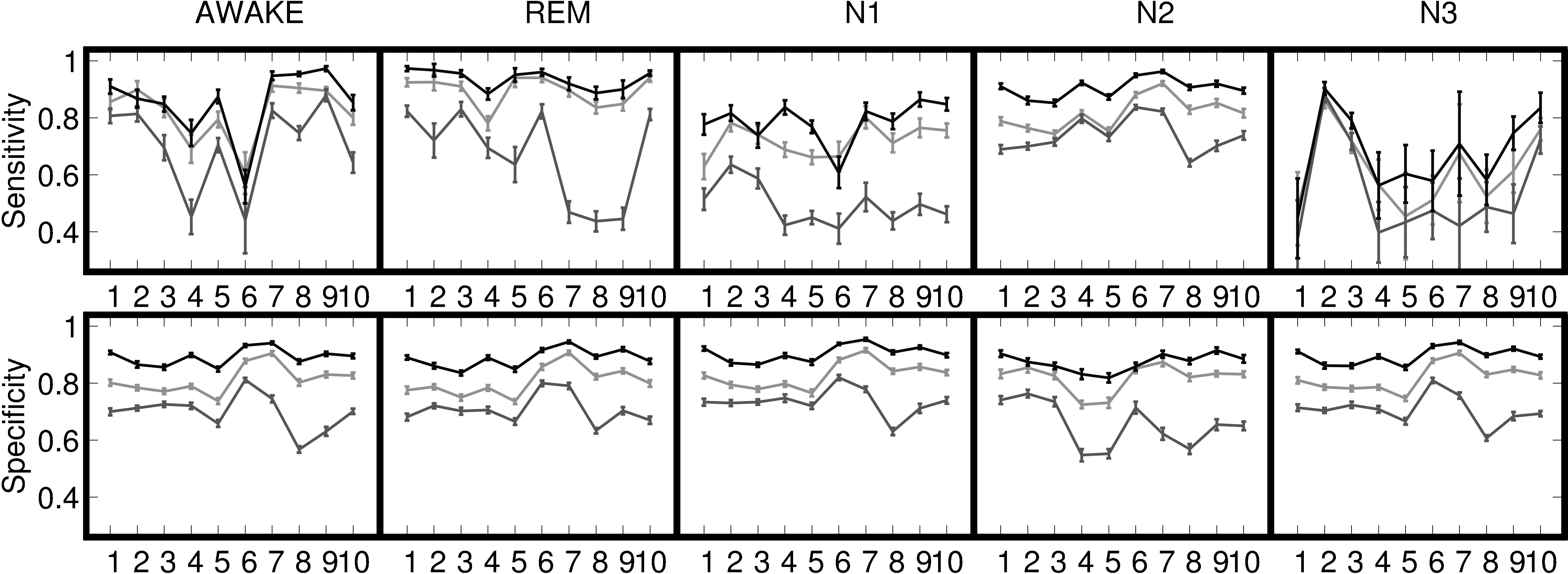} 
\end{centering}
\caption{The error bar of the performance of each features for predicting the sleep stage. The upper (resp. lower) subfigure is the sensitivity (resp. specificity) of predicting different sleep stages by different indices over $25$ repeated random sub-sampling validation. The black (resp. light gray and dark gray) curve is for the Sleep Index (resp. Respiratory Index and EEG Index). The x-axis is the subject index ranging from $1$ to $10$.}
\label{fig:errorbar}
\end{figure*}

With the above preparation, first, we show that the proposed features capturing the sleep information hidden inside the respiratory signal are not only theoretically rigorously supported, but also useful in practice. The overall AC is $81.7\%$. 
The error bar of SE and SP of correlating the Respiratory Index and the sleep stages $\mathcal{R}$ over $25$ repeated random sub-sampling validation for the $10$ subjects is shown in the light gray curve in Figure \ref{fig:errorbar}. The average SE's (resp. SP's) over $10$ subjects for the awake, REM, N1, N2 and N3 stages are $82\%$, $89\%$, $72\%$, $82\%$ and $62\%$ (resp. $81\%$, $81\%$, $83\%$, $82\%$ and $82\%$).

Second, we show that the EEG Index also correlates with the sleep stages. The overall AC is $71.6\%$. 
The error bar of the SE and SP over $25$ repeated random sub-sampling validation for the $10$ subjects is shown in the dark gray curve in Figure \ref{fig:errorbar}. The average SE's (resp. SP's) over $10$ subjects for the awake, REM, N1, N2 and N3 stages are $70\%$, $67\%$, $50\%$, $74\%$ and $54\%$ (resp. $70\%$, $71\%$, $73\%$, $65\%$ and $71\%$).

Next, we combine all the features extracted from the respiratory signal and the EEG signals and show the result is better than simply using the Respiratory Indices or EEG Indices. The overall AC is $89.3\%$. The error bar of the SE and SP over $25$ repeated random sub-sampling validation for the $10$ subjects is shown in the black curve in Figure \ref{fig:errorbar}. The average SE's (resp. SP's) over $10$ subjects for the awake, REM, N1, N2 and N3 stages are $85\%$, $94\%$, $79\%$, $90\%$ and $68\%$ (resp. $89\%$ $89\%$, $90\%$, $87\%$ and $90\%$).

We then apply the Mann-Whitney U test to test if the Sleep Index better predicts sleep stage than the Respiratory Index under our setup. The p value less than $0.01$ is considered significant. For the $25$ realizations of sub-sampling validation from $10$ subjects, we obtained $250$ SE's and $250$ SP's for different indices respectively. The Mann-Whitney U test is applied to see if the SE's and SP's are significantly different. The performance of the Sleep Index compared with the Respiratory Index on the awake, REM, N1, N2 and N3 stages in the sense of SE (resp. SP) are all significant with p-values $<0.001$ ($<0.001$).

\begin{figure*}[t]
\begin{centering}
\includegraphics[width=.7\textwidth]{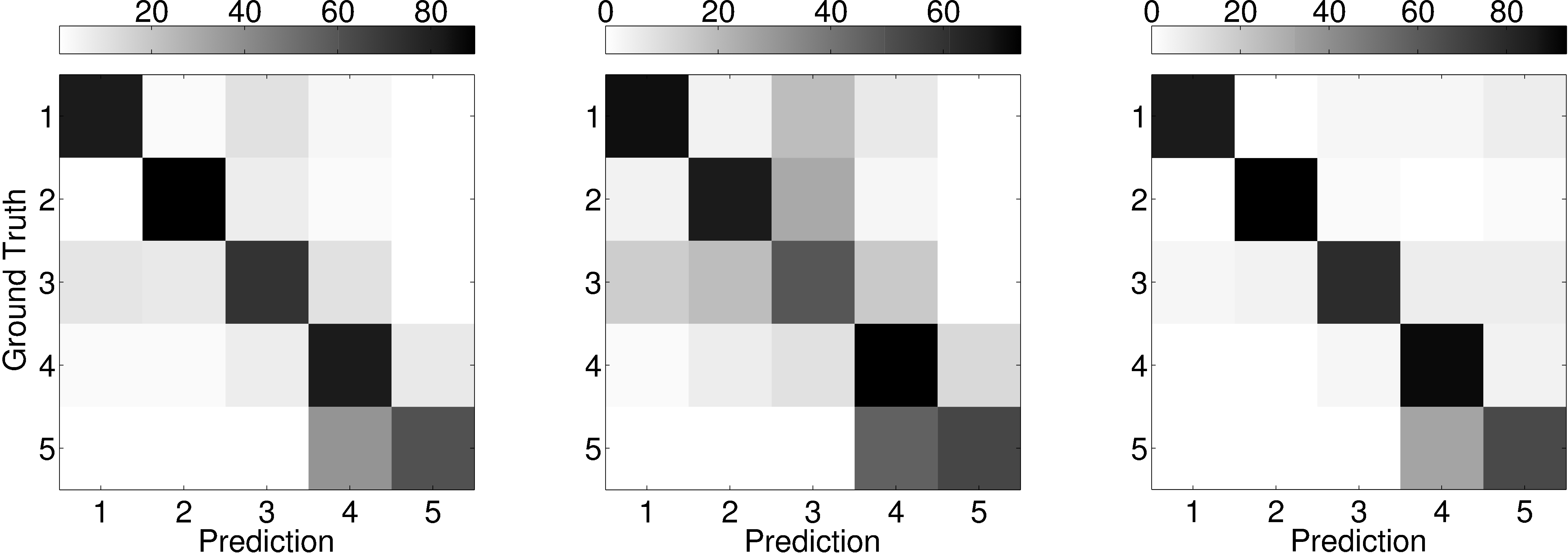} 
\end{centering}
\caption{The averaged confusion percentage matrices over all subjects and sub-sampling realizations based on the Respiratory Index (resp. EEG Index and Sleep Index) is shown in the left (resp. middle and right) subfigure. The percentage is represented by the color. The darker the entry is, the higher the value is. The precise value is shown in the color bar on the top of each matrix. Here, \texttt{1} (resp. \texttt{2}, \texttt{3}, \texttt{4} and \texttt{5}) in the x- and y-axis tick label stands for awake (resp. REM, N1, N2 and N3). It is clear to see the inclination of mis-classifying N3 into N2.}
\label{fig:confusion}
\end{figure*}

Lastly, to better present the classification result, the averaged confusion percentage matrices over all subjects and sub-sampling realizations based on the Respiratory Index, EEG Index and the Sleep Index are shown in Figure \ref{fig:confusion}. Note that the diagonal entries are the SE's of sleep stage prediction. 

\section{Discussion}\label{Section:Discussion}

The results in Section \ref{Section:Result} show that an accurate estimation of all sleep stages by solely analyzing the respiratory signal is possible by combining EIG and SST. Indeed,  in addition to the overall AC $81.7\%$, the average SE is greater than $72\%$ except N3, and the average SP is greater than $81\%$. On the other hand, we mention that while the features of the respiratory signal extracted by EIG and SST are complementary, only EIG can be applied to the EEG signal, since the EEG signal can not be modeled by the adaptive harmonic model. 
The overall AC based on EIG applied to the EEG signals is $71.8\%$, the SE is greater than $67\%$, except N1 and N3, and the average SP is greater than $65\%$.
Namely, the performance of the EEG Index is not better than the Respiratory Index. Nevertheless, we see an improvement in the classification performance based on the Sleep Index, which contains information from both the respiratory and EEG signals. The overall AC is increased to $89.3\%$, the average SE is now greater than $79\%$ except N3, and the average SP is greater than $87\%$. In addition, it has been shown that the SE and SP of the Sleep Index are significantly better than those of the Respiratory Index. Moreover, the confusion percentage matrices also indicate that except N3, the mis-classification does not land in any specific sleep stage. The above findings lead to the following two tentative conclusions: 1. in addition to the EEG signals, the respiratory signal contains ample information about the sleep stage; 2.  combining the relevant but different information hidden inside the respiratory and the EEG signals leads to a better result.

The main innovation in our sleep depth analysis is the combination of the clinical observation and modern adaptive signal processing techniques. From the clinical standpoint, we take the well known physiological fact that in addition to the brain activity, sleep is a global dynamical process involving different sub-system dynamics, in particular the significant changes in the respiratory pattern among different sleep stages. From the signal processing standpoint, we emphasize the importance of the nonlinearity controlling the sleep cycle and focus on finding suitable mathematical tools not only adaptive to the signal but also with sufficient rigorousness to quantify the clinical observation. Indeed, since the unaccessible intrinsic sleep dynamics is reflected in the nonlinear behavior of the respiration, and the two modern signal processing techniques, EIG and SST, have being theoretically studied to well quantify these nonlinearity,
we obtain effective features by analyzing the recorded respiratory signal, which surrogate the intrinsic sleep dynamics.

The meaning of accuracy deserves some discussion. It is well known that the sleep stage determination agreement between different sleep experts is limited to $85\%$ even when the subjects under examination are normal, and it is even worse on the abnormal subjects \cite{Norman_Pal_Stewart_Walsleben_Rapoport:2000}\footnote{It is reported in \cite{Norman_Pal_Stewart_Walsleben_Rapoport:2000} that the mean agreement in the normal subset is higher (mean 76\%, range 65-85\%) than in the subset of sleep disordered breathing (mean 71\%, range 65-78\%).}. Although our cases are not diagnosed as sleep apnea, they cannot be considered as in the normal population either, thereby attaining accuracy rates higher than $80\%$ in our subjects may not be meaningful. 
On the other hand, we found that the classification of N3 stage is consistently worse and its mis-classification tends to land in N2, as is shown in Figure \ref{fig:confusion}. Notice that the subjects in our study are on average $48$ years old, and the distribution of N3 sleep stages in the normal population of this age is $4-20\%$. However, the N3 sleep stages in our study cases is $3.1\%\pm 3.26\%$ with $25\%$ and $75\%$ quantiles $0.8\%$ and $5\%$ respectively, which is much fewer than those in the normal population. Since the number of N3 in the training set is relatively small, even by applying the weighted SVM to handle the unbalanced data, we do not expect to attain a compatible classification rate of N3. This unbalanced training set issue, combined with the stable breathing pattern during N2 and N3, might explain the inclination of mis-classifying N3 into N2. 
Furthermore, while the accuracy of our classification is compatible with/better than the state-or-art reported results, we are able better classify between different sleep stages. Indeed, in \cite{Guerrero-Mora_Elvia_Bianchi_Kortelainen:2012}, the overall accuracy of classifying awake and sleep is $83.6\%$ based on the respiratory signal; in \cite{Chung_Choi_Kim_Lim_Choi_Jeong_Park:2007} an averaged respiratory rate classifies REM and NREM with accuracy over $85\%$; in \cite{Sloboda_Das:2011}, a notch filter based IF estimator is applied to extract respiratory features, which classifies awake, REM and NREM with mean accuracy approximately $70\%$; in \cite{Chen_Cheng_Wu:2013}, the IF estimated by SST is shown to be able to distinguish awake, REM, shallow and N3 with statistical significance.
With the above discussions, we conclude that our features and the selected classifier are accurate. 

The sleep depth estimation by the EEG Index is inferior with respect to the traditional EEG analysis. To understand this result, we briefly revisit how a sleep expert determines the sleep stage. Based on the protocol criteria, in addition to an EEG signal of duration that exceeds 30 seconds, the expert also takes into account past and future EEG signals to determine the sleep stage. However, in our study, the EEG Index is based on the signal in epochs of length $2.56$ seconds. The choice of $2.56$-second interval is for the sake of balancing between the dimension and number of data points in EIG. Although the local covariance structure of the EEG signal is taken into account in the EIG analysis, this relationship is different from the protocol criteria. As a result, we do not expect to obtain a compatible stratification power. 
However, we see that even if we only focus on these short-term EEG signals, we still can predict the sleep stage up to some accuracy and it does help to attain a better classification rate when combined with the Respiratory Index. This hints the possibility that some useful information is hidden inside a finer scale EEG signals. This interesting potential will be reported in the future study. 

The discussion would not be complete without mentioning the shortcomings of our study. First, we focus on a small database containing only $10$ relatively normal subjects in this study. To confirm the usefulness of the proposed features, we need to study a larger database with different types of subjects. Second, the chosen features, in particular the features selected by EIG, are subject-dependent. Indeed, different subjects might have different dynamical systems and the number of dominant components determined by EIG might vary. A theoretical and practical study of integrating the proposed features among different subjects is undergoing. 

In conclusion, by applying modern signal processing techniques to EEG and respiratory signals, we find a set of suitable features, which allow us to predict the sleep stages accurately. In addition to gaining insight into the dynamics controlling the sleep dynamics, the automatic annotation system based on the analysis might lead to an objective classification as well as reduce the required human expert analysis involved in sleep evaluation.

\section*{Acknowledgements} 
Hau-tieng Wu and Ronen Talmon thank the helpful discussions with Professor Ronald Coifman. Ronen Talmon acknowledges the support by the European Union's - Seventh Framework Programme (FP7) under Marie Curie Grant No. 630657.  Yu-Lun Lo acknowledges the support by Taiwan National Science Council grant 101-2220-E-182A-001 and 102-2220-E-182A-001. 

\bibliographystyle{IEEEtran} 
\bibliography{sleep}

\end{document}